\documentclass[fleqn,twoside]{article}
\usepackage{espcrc2}
\usepackage[dvips]{graphicx,graphics,color,epsfig}
\newcommand{\klll}    {\mbox{$K^\circ_L \! \rightarrow \! \ell^+ \ell^-$ }}
\newcommand{\klgg}    {\mbox{$K^\circ_L \! \rightarrow \!  \gamma\gamma$ }}
\newcommand{\klmm}    {\mbox{$K^\circ_L \! \rightarrow \! \mu^+ \mu^-$ }}
\newcommand{\klee}    {\mbox{$K^\circ_L \! \rightarrow \! e^+ e^-$ }}
\newcommand{\klme}    {\mbox{$K^\circ_L \! \rightarrow \! \mu e$ }}
\newcommand{\klpnn}   {\mbox{$K^\circ_L \! \rightarrow \! \pi^\circ \nu \overline{\nu}$ }}
\newcommand{\klmmee}  {\mbox{$K^\circ_L \! \rightarrow \! \mu^+ \mu^- e^+ e^-$ }}
\newcommand{\klpopo}  {\mbox{$K^\circ_L \! \rightarrow \! \pi^\circ \pi^\circ$ }}

\newcommand{\kzpnn}    {\mbox{$K \! \rightarrow \! \pi \nu \overline{\nu}$ }}

\newcommand{\kmng}    {\mbox{$K^+ \! \rightarrow \! \mu^+ \nu_\mu \gamma$ }}

\newcommand{\kpx}     {\mbox{$K^+ \! \rightarrow \! \pi^+ X^\circ$ }}
\newcommand{\kpp}     {\mbox{$K^+ \! \rightarrow \! \pi^+ \pi^\circ$ }}
\newcommand{\kppg}    {\mbox{$K^+ \! \rightarrow \! \pi^+ \pi^\circ \gamma$ }}
\newcommand{\kpen}    {\mbox{$K^+ \! \rightarrow \! \pi^\circ e^+ \nu_e$ }}

\newcommand{\kppnn}   {\mbox{$K \! \rightarrow \! \pi^+ \pi^\circ \nu \overline{\nu}$ }}
\newcommand{\kppen}   {\mbox{$K^+ \! \rightarrow \! \pi^+ \pi^- e^+ \nu_e$ }}

\newcommand{\kpnn}    {\mbox{$K^+ \! \rightarrow \! \pi^+ \nu \overline{\nu}$ }}
\newcommand{\kpg}     {\mbox{$K^+ \! \rightarrow \! \pi^+ \gamma$ }}
\newcommand{\kpgg}    {\mbox{$K^+ \! \rightarrow \! \pi^+ \gamma \gamma$ }}
\newcommand{\kpmm}    {\mbox{$K^+ \! \rightarrow \! \pi^+ \mu^+ \mu^-$ }}
\newcommand{\kpee}    {\mbox{$K^+ \! \rightarrow \! \pi^+ e^+ e^-$ }}

\newcommand{\kpme}    {\mbox{$K^+ \! \rightarrow \! \pi^+ \mu^+ e^-$ }}

\newcommand{\kenmm}   {\mbox{$K^+ \! \rightarrow \! e^+ \nu \mu^+ \mu^-$ }}

\newcommand{\pnn}     {\mbox{$\pi^+ \nu \overline{\nu}$ }}

\newcommand{\bsmix}   {\mbox{$B_s^\circ$---$\overline{B_s^\circ}$ }}
\newcommand{\bsbd}    {\mbox{$\Delta M_{B_d}/\Delta M_{B_s}$ }}
\newcommand{\bpsiks}  {\mbox{$B^\circ_d \! \rightarrow \! \psi K^\circ_S$ }}
\newcommand{\vtd}     {\mbox{$V_{td}$ }}
\newcommand{\Vtd}     {\mbox{$| V_{td} |$ }}

\title{
\vspace{-1.0cm}
\rightline{\small\rm BNL--69052}
\vspace{-0.25cm}
\rightline{\small\rm March 1, 2002}
\vspace{1cm}
Kaon Physics at BNL \\
{\small To be published in the
{\it Proceedings of the  The Fifth KEK Topical Conference
                        - Frontiers in Flavor Physics;
KEK, Tsukuba, Japan; November 20-22, 2001};
S.~Hashimoto and T.K.~Komatsubara, Eds.  }
 }

\author{S.H. Kettell\address{Brookhaven National Laboratory\\
	Upton, NY 11973-5000}}
       
\begin{document}

\begin{abstract}
The rare kaon decay program at BNL is summarized. A brief review of
recent results is provided along with a discussion of prospects for
the future of this program.  The primary focus is the two golden
modes: \kpnn and \klpnn.  The first step in an ambitious program to
precisely measure both branching ratios has been successfully
completed with the observation of two \kpnn events by E787. The E949
experiment is poised to reach an order of magnitude further in
sensitivity and to observe $\sim$10 Standard Model events.

\vspace{1pc}
\end{abstract}

\maketitle

\section{INTRODUCTION}

The AGS has had a broad and rich program in kaon physics.  With the
recent successful commissioning of the Relativistic Heavy Ion Collider
(RHIC), the primary role of the AGS has shifted to become an injector
of heavy ions for RHIC. Nevertheless, the AGS remains the highest
intensity proton synchrotron in the world and is designed to be
available for $\sim$20 hours/day when not filling RHIC, and as such
retains an important role in the US high energy physics (HEP) program.
DOE and BNL have approved and agreed to fund one new HEP experiment to
run at the AGS between RHIC fills: the E949 experiment which seeks to
make a precise measurement of the branching ratio B(\kpnn).

\subsection{\boldmath \kzpnn}

The kaon physics program at the AGS is centered on the two golden
modes: \kpnn and \klpnn. These modes are interesting as there is
essentially no theoretical ambiguity in extracting
fundamental CKM parameters from measurements of the branching ratios\cite{sd_kpnn,bb3}.
The theoretical uncertainty in B(\kpnn) is $\sim$7\% and is even
smaller in B(\klpnn), only $\sim$2\%; in both cases the hadronic matrix
element is extracted from the \kpen ($K_{e3}$) decay rate.

The unitarity of the CKM matrix can be expressed as
\[V^*_{us}V_{ud} + V^*_{cs}V_{cd} + V^*_{ts}V_{td} 
= \lambda_u + \lambda_c + \lambda_t = 0 \nonumber
\]
with the three vectors $\lambda_i\equiv V^*_{is}V_{id}$ converging to
form a very elongated triangle in the complex plane. The length of
first vector $\lambda_u=V^*_{us}V_{ud}$ is precisely determined from 
$K_{e3}$ decay.  The height of the triangle, $Im\lambda_t$,
can be measured by \klpnn and the length of the third side
$\lambda_t=V^*_{ts}V_{td}$ is measured by \kpnn.  Measurements of the
two \kzpnn modes, along with the well known $K_{e3}$, will completely determine
the unitarity triangle.

Comparison of CKM parameters as measured from the golden 
\kzpnn, \bpsiks and \bsbd modes, provide the best
opportunity to over-constrain the unitary triangle and to search for
new physics. In particular, comparisons of
\begin{itemize}
  \item $|\vtd|$ from \kpnn and from the ratio
of the mixing frequencies of $B_d$ and $B_s$ mesons \bsbd~\cite{bb3}, 
  \item $\beta$ from B(\klpnn)/B(\kpnn) and from the time dependent asymmetry in the decay \bpsiks~\cite{sinb,gr-nir}
\end{itemize}
offer outstanding opportunities to explore the Standard Model (SM) picture of CP-violation.

The SM prediction for the \kzpnn branching ratios are B(\kpnn) =
$(0.75 \pm 0.29) \times 10^{-10}$ and B(\klpnn) = $(0.26 \pm 0.12)
\times 10^{-10}$~\cite{sd_kpnn}. In addition, an upper limit on B(\kpnn)
can be derived in a theoretically unambiguous way from the current limit
on $B_s$ mixing; this limit is B(\kpnn) $< 1.32 \times 10^{-10}$~\cite{damb_isidori}.

\section{SUMMARY OF RECENT RESULTS}

A large number of new results are available from several recently
completed rare kaon decay experiments at BNL (running during
1995--98). There were three experiments running during this period ---
E787: Search for \kpnn, E865: Search for \kpme and E871: Search for
\klme.

\subsection{Lepton Flavor Violation Searches}

Two of these experiments, E865 and E871, made use of the tremendous
kaon flux available at the AGS to push the sensitivity of searches for
physics beyond the SM (BSM), in particular, to search for lepton flavor
violating decays. These experiments have, or will soon, reach single
event sensitivities of $\sim$$10^{-12}$ and have excluded or severely
limited many possible extensions to the SM. A summary of limits from
these searches for BSM physics is provided in
Table~\ref{table:BSM}. The E871 experiment is finished with all data
analysis, but new results on \kpme and other modes are expected from
E865 in the near future.
\begin{table}[htb]
\caption{Searches for decays not allowed in the standard model}
\label{table:BSM}
\begin{tabular}{lll} \hline
Decay Mode  & \multicolumn{1}{c}{BR} & Reference \\ \hline
\klme  & $<4.7\times10^{-12}$ &E871-98~\cite{prl_klme} \\ 
\kpme  & $<2.8\times10^{-11}$ &E865-00~\cite{prl_kpme} \\ 
\kpx   & $<5.9\times10^{-11}$ & E787-02~\cite{prl_kpnn} \\ 
$K^+\rightarrow\pi^-\mu^+ e^+$ & $<5.0\times10^{-10}$ &E865-00~\cite{prl_kopp} \\ 
$K^+\rightarrow\pi^+\mu^- e^+$ & $<5.2\times10^{-10}$ & E865-00~\cite{prl_kopp} \\
$K^+\rightarrow\pi^- e^+ e^+$ & $<6.4\times10^{-10}$ & E865-00~\cite{prl_kopp} \\ 
$K^+\rightarrow\pi^-\mu^+\mu^+$ & $<3.0\times10^{-9}$ & E865-00~\cite{prl_kopp} \\
\kpg   & $< 3.6\times10^{-7}$ & E787-02~\cite{prd_kpg} \\
\end{tabular}
\end{table}

\subsection{Chiral Perturbation Theory}

A number of other modes have been observed for the first time, or with
significantly larger statistics than was previously available. These
modes are of interest as a testing ground for Chiral
Perturbation Theory (ChPT). A summary of these measurements is
provided in Table~\ref{table:ChPT}.
\begin{table*}[htb]
\caption{Modes of interest to Chiral Perturbation theory}
\label{table:ChPT}
\begin{tabular}{llrll} \hline
Decay Mode  &  \multicolumn{1}{c}{Branching Ratio} & Events& Experiment  & Ref \\ \hline
\klmm & $(7.24\pm0.17)\times10^{-9}$ & 6200 & E871 (2000) & PRL {\bf 84}:1389~\cite{prl_klmm}\\ 
\klee & $(8.7^{+5.7}_{-4.1})\times10^{-12}$& 4 & E871 (1998) & PRL {\bf 81}:4309~\cite{prl_klee}\\
\kpgg &  $(6.0\pm1.5\pm.7)\times10^{-7}$& 26 & E787 (1997) & PRL {\bf 79}:4079~\cite{prl_kpgg} \\ 
\kpee&$(2.94\pm.05\pm.13)\times10^{-7}$&10,300&E865 (1999) & PRL {\bf 83}:4482~\cite{prl_kpee} \\ 
\kpmm&$(5.0\pm.4\pm.6)\times10^{-8}$&200&E787 (1997) & PRL {\bf 79}:4756~\cite{prl_kpmm} \\ 
\kpmm&$(9.22\pm.60\pm.49)\times10^{-8}$&430&E865 (2000) & PRL {\bf 84}:2580~\cite{prl_kpmm_787} \\ 
\kppg(DE) &$(4.72\pm.77\pm.28)\times10^{-6}$ & 360 & E787 (2000) & PRL {\bf 85}:4856~\cite{prl_kppg} \\ 
\kppen   & $(4.11\pm.01\pm.11)\times10^{-5}$  & 388,270 & E865 (2001) & PRL {\bf 87}:221801~\cite{prl_kppen}\\
\kmng(SD)& $(1.33\pm.12\pm.18)\times10^{-5}$ & 2588 & E787 (2000) & PRL {\bf 85}:2256~\cite{prl_kmng} \\
\kenmm   & $< 5.0\times10^{-7}$ & 0 & E787 (1998) & PR {\bf D58}:012003~\cite{prl_kenmm} \\ 
\end{tabular}
\end{table*}

A byproduct of the E871 search for \klme, and a demonstration of the
capabilities of the experiment is the observation of the \klee
decay, with
the smallest
BR ever measured for an elementary particle decay. Along with \klee, a
very high statistics measurement of the until recently considered
`rare' decay \klmm was made.  This mode is very interesting due to its
historical significance, and to the fact that the short distance
component of this decay is sensitive to $\rho$ or $Re\lambda_t$. The
dominant contribution to this decay is from two real photons and is
well known from QED and the measured value of B(\klgg).  It may be
possible, with additional theoretical work~\cite{chpt_klmm1} and/or
new data (for example, from \klmmee from KTeV at FNAL) to determine the long
distance dispersive contribution to this mode, although there is some
controversy on this point~\cite{chpt_klmm2}.  If all long distance
contributions can be calculated, this measurement will be able to
provide an independent constraint on $\lambda_t$. A
summary of the \klll modes can be found in Table~\ref{table:ChPT}.

\subsection{CKM Matrix and CP-violation}

The third experiment, E787, was designed to search for the decay
\kpnn.  This experiment made use of the tremendous flux of low energy
kaons to measure this very rare mode (branching ratio $\sim10^{-10}$)
with only one detectable particle in the final state.  A summary of
\pnn results is provided in Table~\ref{table:CKM}.
\begin{table*}[htb]
\caption{Measurements of the CKM Matrix and CP-violation}
\label{table:CKM}
\begin{tabular}{llrll} \hline
Decay Mode  &  \multicolumn{1}{c}{Branching Ratio} & Events& Experiment  & Ref \\ \hline
\kpnn&$(1.57^{+1.75}_{-0.82})\times10^{-10}$&2&E787 (2002) & PRL {\bf 88}:041803~\cite{prl_kpnn} \\ 
\klpnn($\pi^+\nu\overline{\nu}$) &$<1.7\times 10^{-9}$& ---  & E787 (2002) & indirect~\cite{prl_kpnn} \\ 
\kppnn &$<4.3\times 10^{-5}$& 0  & E787 (2001) & PR {\bf D63}:032004~\cite{prd_kppnn} \\  
\end{tabular}
\end{table*}

\begin{figure}[htb]
\psfig{file=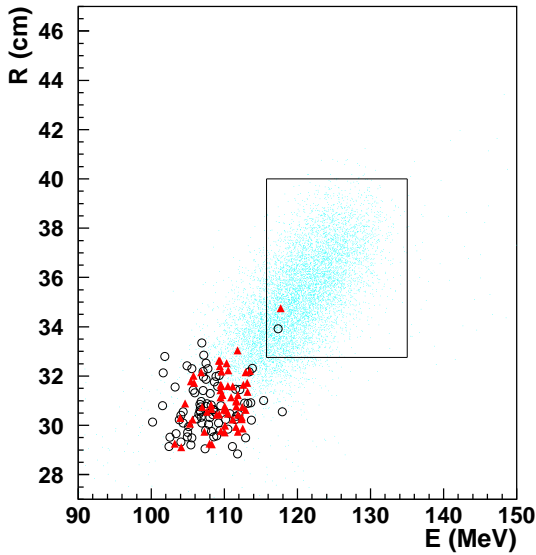,width=2.75in,angle=0}
\caption{Final E787 plot of range vs. energy for events passing all
other cuts. The circles are for 1998 data and the triangles are for
1995--97 data. Two clean \kpnn events are seen in the box. The
remaining events are \kpp background. A \kpnn Monte Carlo sample
(dots) is overlayed on the data.}
\label{fig:pnn}
\end{figure}

The first \kpnn signal was observed in the 1995 data sample of the
E787 experiment~\cite{prl_kpnn_95}. No new events were seen in the
data sample from 1996--97~\cite{prl_kpnn_97}, and with a background of
$0.08\pm0.02$ events and a signal of one event a branching ratio of
$1.5^{+3.4}_{-1.2}\times 10^{-10}$ was measured. That event was in
fact in a very clean region of the signal box with a standard model
signal to background ratio of 35. An analysis of the final E787 data
sample from the last run in 1998 has recently been
reported~\cite{prl_kpnn}. With a measured background of
$0.066^{+0.044}_{-0.025}$, one new event was observed.  The final plot
of range vs. energy from the combined E787 1995--98 data sample for
events passing all other cuts is shown in Figure~\ref{fig:pnn}.
The branching ratio, as determined from these two events is
$1.57^{+1.75}_{-0.82}\times 10^{-10}$. This branching ratio is a
factor of two higher than expected in the SM and is higher than
allowed by the current limit on $B_s$ mixing. Of course, the
uncertainty on the BR measurement is large due to limited
statistics and new data from the E949 experiment are eagerly awaited.

\begin{figure}[htb]
\psfig{file=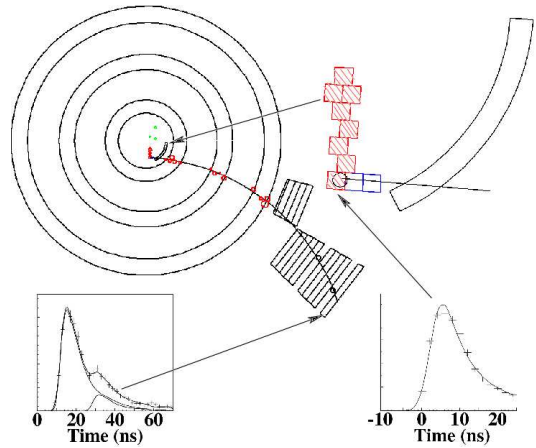,width=2.75in,angle=0}
\caption{Event display for the second \kpnn event, discovered in the
E787 data sample from 1998. The upper left shows an end view of the
detector. The top right shows an expanded view of the target region,
with a view of the digitized pulse in the fiber where the kaon stopped
in the lower right. The lower left shows the digitized
$\pi^+\longrightarrow\mu^+$ decay signal in the scintillator where the
pion stopped.}
\label{fig:event}
\end{figure}

The new event found in the 1998 data sample is in a relatively clean
region of the accepted signal region: the SM signal to background
ratio for this event is 3.6. An event display is shown in
Figure~\ref{fig:event}.

From the 90\% CL limits 
$ 0.56\times 10^{-10}<B(\kpnn) <3.9 \times 10^{-10} $
a limit on the branching ratio of the neutral mode can be
derived~\cite{gr-nir}
\begin{eqnarray*}
B(\klpnn) & < & 4.4\times B(\kpnn) \\ \nonumber
 & < & 1.7\times10^{-9} \hspace{0.3cm} {\rm (90\% \hspace{0.5mm} CL)}.  \nonumber
\end{eqnarray*}
Limits on \Vtd and $\lambda_t$ can be obtained (these are 1-$\sigma$ limits
except for $Im\lambda_t$ which is 90\% CL), 
\begin{eqnarray*}
0.007< &|V_{td}|& <0.030, \\  \nonumber
2.9 \times 10^{-4} < &|\lambda_t|& <1.2 \times10^{-3}, \\  \nonumber
-0.88 \times 10^{-3} < &Re\lambda_t& < 1.2 \times 10^{-3},\\  \nonumber
&Im\lambda_t& < 1.1 \times10^{-3}.  \nonumber
\end{eqnarray*}

\begin{figure}[htb]
\psfig{file=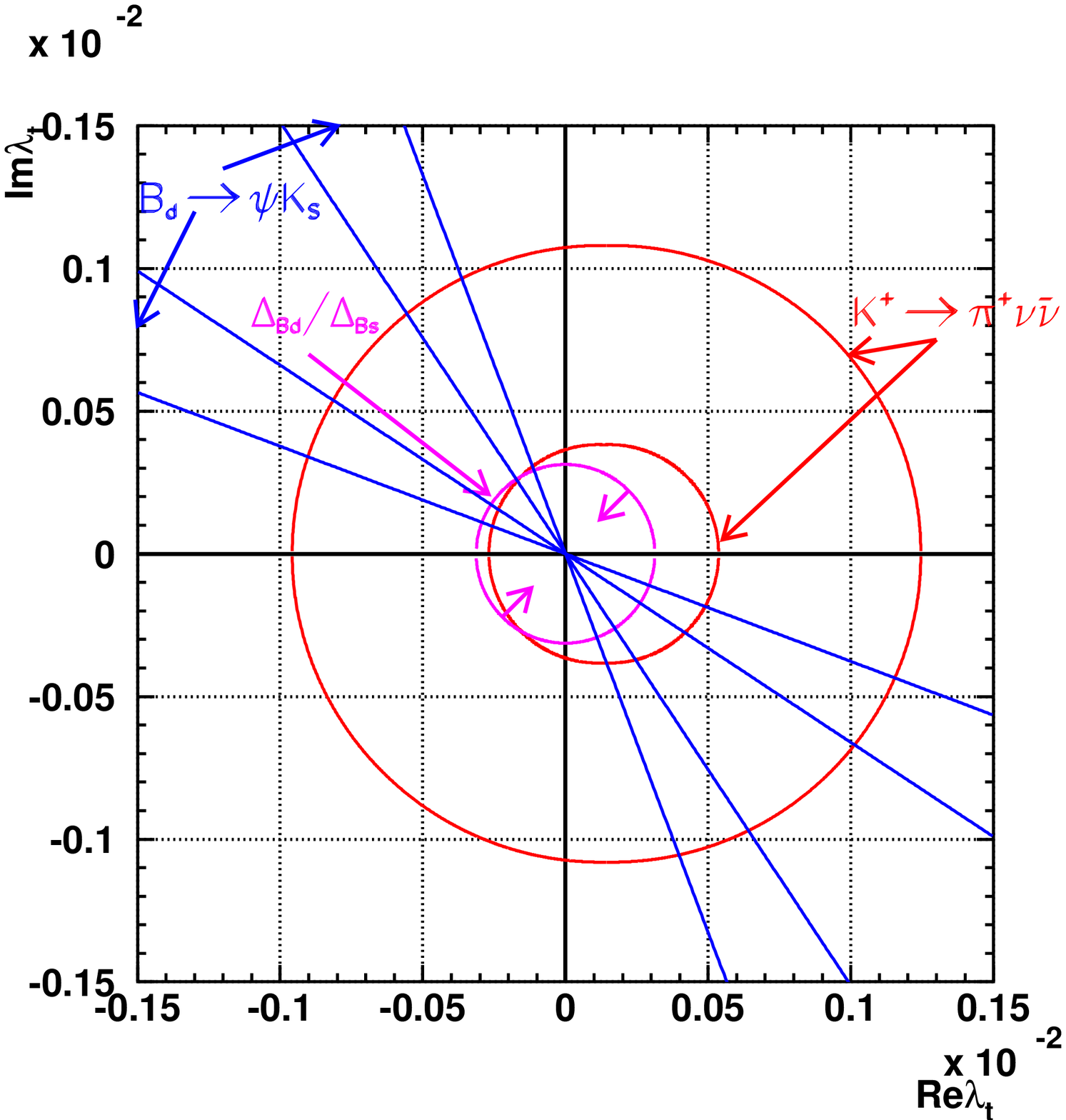,width=3.0in,angle=0}
\caption{Constraints on $\lambda_t$ from B(\kpnn),
\bsbd and \bpsiks. The experimental measurements for \bsbd and \kpnn
are 90\% CL limits and for \bsbd is a 95\% CL limit.
The theoretical uncertainties in all of these modes are small.  A
measurement of \klpnn will provide a constraint on $Im\lambda_t$.}
\label{fig:rho_eta}
\end{figure}

Even with the large statistical error, this new measurement of
B(\kpnn) provides a non-trivial contribution to global fits of the CKM
parameters~\cite{damb_isidori}.  The constraints on $\lambda_t$ from
this result are shown in Figure~\ref{fig:rho_eta}. The constraints
from the other golden B modes, \bsbd and \bpsiks are shown on the same
plot. One can immediately see that the allowed region is tightly constrained
to a narrow crescent by \kpnn and \bsbd.
New data from the successor to E787, E949 will make a significant
contribution to our knowledge of the CKM parameters.

In addition, E787 has searched for the decay \kpnn in the pion
kinematic region below the \kpp ($K_{\pi2}$)
peak~\cite{prl_kpnn2}. This region contains more of the \kpnn phase
space, but is complicated by a significant background from \kpp decays
with the $\pi^+$ scattering in the scintillating fiber target and
down-shifting its kinematics into the search region. The data from the
1996 run of E787 has been analyzed ($\sim$20\% of the entire E787 data
sample). One event was observed in the search region, consistent with
the background estimate of $0.73\pm 0.18$. This implies an upper limit
on B(\kpnn) $< 4.2\times 10^{-9}$ (90\% C.L.), and is consistent with
the 2 events observed above the $K_{\pi2}$ peak and the SM spectrum. Some additional
reduction of the background levels in the remaining E787 data may be
possible, but the major focus will shift to the new E949 experiment,
which has significantly enhanced photon veto capabilities that will further
suppress this background. In addition, the next experiment after
E949, CKM at FNAL, will be essentially free of this background
since there is no stopping target.

\section{CURRENT PROGRAM}

The current high energy physics program at BNL consists of one
experiment: E949 --- A measurement of the branching ratio
B(\kpnn). This is the first of the AGS-2000 experiments to be approved
and is to be funded by DOE for 6000 hours of running beginning in
2002. Unfortunately, the FY02 run will fall substantially short of
the planned running time, so the expectation is that E949 will need to
run beyond FY03.

\subsection{E949}

E949 is an upgraded version of the E787 experiment, planning to
capitalize on the full AGS beam to collect \kpnn data at 14 times the
rate of the E787 run in 1995. The new detector has substantially
upgraded photon veto capabilities, enhanced tracking, triggering,
monitoring, and DAQ capability, and will run at a higher AGS duty
factor and a lower kaon momentum (with an increased fraction of
useful, stopped kaons). It has been designed to reach a sensitivity of
at least 5 times beyond E787 and observe of 5--10 SM events. The
background level for E949 measurement of B(\kpnn) above the $K_{\pi2}$
peak is reliably projected from E787 data to be $\sim$10\% of the
Standard Model signal.

\begin{figure}[htb]
\psfig{file=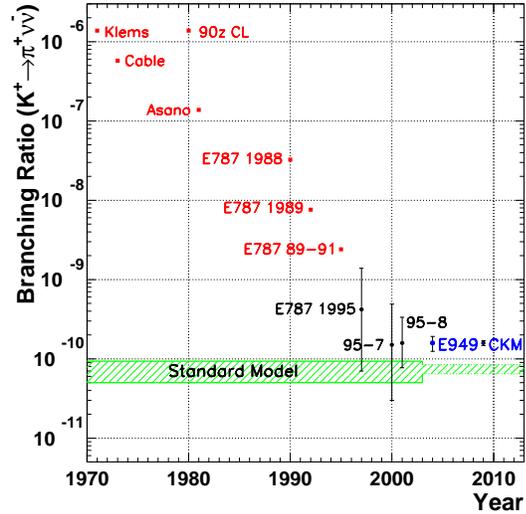,width=3.0in,angle=0}
\caption{History of the search for \kpnn. The squares represent 90\%
CL limits, the dark circles are the E787 observation of \kpnn, and the
projections of the current BR to the proposed E949 and CKM
sensitivities. The prediction from the standard model is expected to
narrow considerably once \bsmix mixing has been observed.
}
\label{fig:hist}
\end{figure}

E949 should see up to 10 SM events (or 20 events at the branching
ratio measured by E787) within the next couple of years. This is an
exciting opportunity to make a significant contribution to quark
mixing and CP-violation that should be fully exploited.  A history of
the search for \kpnn is shown in Figure~\ref{fig:hist}.
The next step towards a precision measurement of B(\kpnn) will be the
CKM experiment at FNAL. CKM has been given scientific (Stage--1)
approval by FNAL and could be running by 2007. CKM plans to use a
novel technique for \kpnn: a decay in flight experiment, with
redundant kinematic constraints from a reasonably conventional
momentum spectrometer and a novel velocity spectrometer based on RICH
counters.  CKM expects to observe 100 SM signal events in a two year
run, using Main Injector pulses not needed by the Tevatron. CKM
will require less than 20\% of the flux from the Main Injector, but
will require a slow extracted spill of $\sim$1
second. Figure~\ref{fig:hist} shows a projected measurement of
B(\kpnn) from CKM, assuming the current branching ratio.

\section{FUTURE PLANS}

Other HEP experiments that are under consideration to run at the AGS
include additional running for
measurement of g-2 of the muon, 
a search for muon to electron conversion (MECO) and a search for and
measurement of the \klpnn branching ratio (K0PI0).  The National
Science Board of the National Science Foundation (NSF) has has
approved the construction of the two new large experiments: K0PI0 and
MECO, as components of the Rare Symmetry Violation Proposal
(RSVP). RSVP is planned to be one of the next Major Research Equipment
construction projects at the NSF.

\subsection{K0PI0}

The KOPIO experiment is designed to discover the \klpnn decay and
measure its branching ratio to $\sim$20\%.  KOPIO will make use of a
time-of-flight technique to measure the momentum of the $K_L$ and will
operate at a large targeting angle to improve the $p_K$ resolution and
soften the neutron spectrum to reduce $\pi^\circ$ hadroproduction.
KOPIO will have a substantial photon veto system and make the same
sort of background measurements as E787, directly from the data, with
two independent tools for attacking the major background, \klpopo,
through both kinematics and photon veto. K0PI0 expects to observe 50
SM events, with a background of 50\%. This will allow a determination
of $Im\lambda_t$ to 10\%. K0PI0 is expected to start data collection
in $\sim$2006.

\section{CONCLUSION}

The next decade will be an exciting time for improved understanding of
CP-violation and quark mixing. It is quite likely that precise
measurements of all four ``Golden Modes'' will be made: \bpsiks at the
B-factories, \bsbd most likely at the Tevatron, and \kzpnn at BNL, KEK and
FNAL. These measurements will allow a precise determination of CKM parameters
and provide a critical test of the SM
picture of CP-violation.


\begin{thebibliography}{99}
\bibitem{sd_kpnn} A.Buras, hep-ph/0101336 (2001); 
A.Buras and R.Fleischer, hep-ph/0104238 (2001).
\bibitem{bb3} G.Bucahalla and A.Buras, NP {\bf B548}, 309 (1999).
\bibitem{sinb} G.Bucahalla and A.Buras, PL {\bf B333}, 221 (1994);
G.Bucahalla and A.Buras, Phys.\ Rev.\ {\bf D54}, 6782 (1996);
Y.Nir and M.P.Worah, PL {\bf B423}, 319 (1998);
S.Bergmann and G.Perez,  hep-ph/0007170.
\bibitem{gr-nir} Y.~Grossman and Y.~Nir, PL {\bf B398}, 163 (1997).
\bibitem{damb_isidori} G.~D'Ambrosio and G.~Isidori, hep-ph/0112135
\bibitem{prl_klme} D.Ambrose, {\it et al.}, PRL {\bf 81}, 5734 (1998).
\bibitem{prl_kpme} R.Appel, {\it et al.}, PRL {\bf 85}, 2450 (2000).
\bibitem{prl_kpnn} S.Adler, {\it et al.}, PRL {\bf 88}, 041803  (2002).
\bibitem{prl_kopp} R.Appel, {\it et al.}, PRL {\bf 85}, 2877 (2000).
\bibitem{prd_kpg} S.Adler, {\it et al.}, PR{\bf D65}, 052009 (2002).
\bibitem{prl_klmm} D.Ambrose, {\it et al.}, PRL {\bf 84}, 1389 (2000).
\bibitem{prl_klee} D.Ambrose, {\it et al.}, PRL {\bf 81}, 4309 (1998).
\bibitem{prl_kpgg} S.Adler, {\it et al.}, PRL {\bf 79}, 4079 (1997).
\bibitem{prl_kpee} R.Appel, {\it et al.}, PRL {\bf 83}, 4482 (1999).
\bibitem{prl_kpmm} H.Ma, {\it et al.}, PRL {\bf 84}, 2580 (2000).
\bibitem{prl_kpmm_787} S.Adler, {\it et al.}, PRL {\bf 79}, 4756 (1997).
\bibitem{prl_kppg} S.Adler, {\it et al.}, PRL {\bf 85}, 4856 (2000).
\bibitem{prl_kppen} S.Pislak, {\it et al.}, PRL {\bf 87}, 221801 (2001).
\bibitem{prl_kmng} S.Adler, {\it et al.}, PRL {\bf 85}, 2256 (2000).
\bibitem{prl_kenmm} S.Adler, {\it et al.}, PR {\bf D58}, 012003 (1998).
\bibitem{chpt_klmm1} G.D'Ambrosio, {\em et. al.}, PL {\bf B423}, 385 (1998);
D.G.Dumm and A.Pich, PRL {\bf 80}, 4633 (1998).
\bibitem{chpt_klmm2} G.Valencia, NP {\bf B517}, 339 (1998);
M.Knecht, {\it et al.}, PRL {\bf 83}, 5230 (1999).
\bibitem{prd_kppnn} S.Adler, {\it et al.}, PR{\bf D63}, 032004 (2001).
\bibitem{prl_kpnn_95} S.Adler, {\it et al.}, PRL {\bf 79}, 2204 (1997).
\bibitem{prl_kpnn_97} S.Adler, {\it et al.}, PRL {\bf 84}, 3768 (2000).
\bibitem{prl_kpnn2} S.Adler, {\it et al.}, hep-ex/0201037
\end{thebibliography}
\end{document}